\documentclass[twocolumn,showpacs,preprintnumbers,amsmath,amssymb,floatfix]{revtex4-1}
\usepackage{color}   %option added by cyw eg \pagecolor{blue} \color{red}
\usepackage{graphicx}% Include figure files
\usepackage{dcolumn}% Align table columns on decimal point
\usepackage{bm}% bold math
\usepackage{slashed}

\begin{document}

\def\bb    #1{\hbox{\boldmath${#1}$}}
\def\slam {{\lambda \!\!\raisebox{-0.2ex}{$\bar {}$}~}}

\title{ Reaction Cross Section in Heavy-Ion Collisions  }

\author{Cheuk-Yin Wong}

\affiliation{Physics Division, Oak Ridge National Laboratory, 
Oak Ridge, TN 37831}

%\date{\today}

\begin{abstract}

Previously a compact formula for total reaction cross section for
heavy-ion collisions as a function of energy was obtained by treating
the angular momentum $l$ as a continuous variable.  The accuracy of
the continuum approximation is assessed and corrections are evaluated.
The accuracy of the compact equation can be improved by a simple
modification, if a higher accuracy is required.  Simple rules to
determine the barrier heights and the penetration probability for the
$l$ partial wave from experimental data are presented, for the
collision of identical or non-identical light nuclei.
\end{abstract}

\pacs{ 25.60.Pj  25.60.Dz  }
                                                                         
\maketitle

\section{Introduction}

Nuclear fusion in heavy-ion collisions is an important process in many
phenomena.  The sub-barrier fusion of light nuclei plays an important
role in the evolution of massive stars, the dynamics of white dwarf
Type Ia supernovae, and explosions on the surface of neutron stars
\cite{Gas05}.  The fusion of heavy nuclei just above the barrier is an
important tool in the production of superheavy nuclei \cite{Hof00}.
The fusion of identical  light nuclei at high energies reveals interesting
effects in isolated high-angular-momentum states of the fused system
\cite{Esb12}.  Nuclear fusion of light nuclei is utilized in
accelerator-based heavy-ion inertial fusion for fusion energy
production \cite{HIF10}.

Previously, a simple analytical expression was obtained for the total
reaction cross section $\sigma_r$ in the collision of nuclei $A_1$ and $A_2$
as a function of the collision energy $E$
\cite{Won73}
\begin{eqnarray}
\sigma_r(E)=\pi R_0^2\frac{\Gamma }{E} \ln\left \{ 1 + \exp\left  (\frac{E-E_0}{\Gamma}\right )\right\},
\label{eq1}
\end{eqnarray}
where $E_0$ is the barrier height for the $l$=0 partial wave,
$\Gamma$=${\hbar\omega}/{2\pi}$ is the energy width parameter in 
the potential barrier penetration probability, $\omega$ is the
frequency of the inverted parabola representing the potential barrier,
and $R_0=r_0(A_1^{1/3}+A_2^{1/3})$ is the spatial separation
between the two nuclei at the potential barrier.  By assuming that the
fusion process is equivalent to the strong-absorption of ingoing waves
passing through the potential barrier, the total reaction cross
section (\ref{eq1}) can be interpreted as the heavy-ion fusion cross
section.

Even though the fusion process involves complicated dynamics of
channel coupling, dynamical distortions, polarizations, deformations,
particle transfers, proximity interactions, and barrier penetrations
\cite{Cit12}, the simple expression of Eq.\ (\ref{eq1}) provides an
efficient way to represent experimental data in terms of important
physical parameters, whose systematics give valuable insights into the
dynamics of the process. The relationship between the barrier
penetration model used in Ref.\ \cite{Won73} and the coupled-channels
calculations for heavy-ion fusion was discussed previously
\cite{Nar86}.  Equation (\ref{eq1}) has been successfully applied to
describe fusion cross sections in the collision of many projectile and
target combinations \cite{Cit12}.

As the range of fusion energy in astrophysical processes and
fusion-energy production extends to the region beyond the sensitivity
of present-day measurements, theoretical extrapolations are needed to
access many relevant rates of fusion reactions \cite{Gas05}.  It is
desirable to examine the accuracy of the simple formula (\ref{eq1}) in
the sub-barrier region.  In the other extreme in higher energy fusion,
the recent interesting observation of the stepwise increase of the
fusion cross section for two identical spin-0 nuclei \cite{Esb12} also
calls for an analysis on  its accuracy in the higher energy regime.  

The simple result of Eq. (1) was obtained by treating the discrete angular
momentum $l$ as a continuous variable.  Such a treatment incurs
errors.  We therefore wish to assess the accuracy of Eq.\ (\ref{eq1})
over the whole energy range and to find how it may be improved if
a higher accuracy is desired.

The simple result of Eq. (1) relies also on the
assumption that the fusion barrier height $E_l$ for the $l$ partial
wave is a linear function of $l(l+1)$.  While such an assumption is a
reasonable concept for many reactions, there are nuclear collisions
for which such an assumption is not valid, as is evidenced by the
disagreement of the ${}^{12}$C+${}^{13}$C data \cite{Not12,Day76} with
the main features of Eq.\ (1).  To diagnose such a pathological case,
it will be useful to develop tools that will enable us to determine
the fusion barrier heights $E_l$ and the penetration probabilities
$P_l(E)$,  as a function of $l$ from experimental data.  Furthermore,
the direct determination of these physical quantities provides useful
insight into the fusion process.  

For those cases where the basic assumption of the linear dependence of
the fusion barrier $E_l$ on $l(l+1)$ does not hold, we would like to
propose alternative methods to describe the reaction cross section.
We wish to design a framework to partition the reaction cross section
such that contributions to different regions of $l$ partial waves can
be singled out for special scrutiny.

This paper is organized as follows.  In Sec. II, we 
evaluate the reaction cross section and develop the
rules for fusion barrier analysis.  In Sec. III, we present
the formulation of the reaction cross section, the continuum
approximation, and its corrections.  In Sec. IV, we give the
numerical results and the comparison with experimental data.  In
Sec. V, we carry out a barrier analysis for
${}^{12}$C+${}^{13}$C and show that the barrier $E_l$ for that
reaction is not a linear function of $l(l+1)$.  We show how the fusion
cross section of such a pathological case can be described by an
alternative method of partitioning the reaction cross section.  We
present the rules for barrier analysis in Sec. VI and the
rules for the penetration probability analysis in Sec. VII, for the
collision of identical or non-identical light nuclei.  In Sec. VIII, we
present our conclusions and discussions.

\section{Reaction Cross section and Fusion Barrier Analysis}

We approximate various barriers for different partial $l$ waves by
inverted harmonic-oscillator potentials of height $E_l$ and frequency
$\omega_l$ in the ingoing- wave strong-absorption model. For a
collision energy $E$, the probability for the absorption of the $l$
partial wave is then given by the Hill-Wheeler penetration probability
formula \cite{Hil53},
\begin{eqnarray}
P_l(E)= \frac{1}{1+\exp \{ 2\pi (E_l -E)/\hbar \omega_l\}}.
\end{eqnarray}•
In consequence, the total reaction cross section as a function of $E$
for the collision of unequal nuclei is \cite{Won73}
\begin{eqnarray}
\sigma_r(E)&=&
\frac{\pi}{k^2} \sum_{l=0,1,2,..} 
\frac{2l+1}{1+\exp \{ 2\pi (E_l -E)/\hbar \omega_l\}}.
\label{eq2}
\end{eqnarray}
The above expression can be cast in a more illuminating form in terms
of the deBroglie wave length $\slam=1/k$.  The natural unit of cross
sectional area in deBroglie wave length scales is $\pi \slam^2$ \cite{Bla52}, 
which can be conveniently called the deBroglie 
cross section.  Using Eq.\ (\ref{eq2}), we can construct the
dimensionless measure of the reaction cross section $\sigma_r$ in
units of $\pi \slam^2$, $({\sigma_r}/{\pi \slam^2})$, at the collision energy $E$, as given by
\begin{eqnarray}
 ( {\sigma_r}/{\pi \slam^2} )(E)&=& \sum_{l=0,1,2,...}f(l)
=f_0+f_1+f_2+f_3+..,~~~
\label{3} 
\end{eqnarray}
where 
\begin{eqnarray}
 f_l=f(l)&=& \frac{2l+1}{1+\exp \{ (E_l - E)/\Gamma_l \}},
\end{eqnarray}
and $\Gamma_l=\hbar \omega_l/ 2\pi$.  We can evaluate $(\sigma_r/\pi
\slam^2)(E)$ at the collision energy $E$.  We find from Eq.\ (\ref{3})
that for the energy $E$ such that
\begin{eqnarray}
l^2 \le (\sigma_r/\pi \slam^2)(E)  \le (l+1)^2,
\label{pc6}
\end{eqnarray}
the reaction cross section at energy $E$ is
\begin{eqnarray}
(\sigma_r/\pi \slam^2)(E)&=&
l^2+\frac{2l+1}{1+\exp\{(E_l-E)/\Gamma\}}+C(l,E),~~~
\label{pc7}
\end{eqnarray}
where the correction term $C(l,E)$ takes into account the width $\Gamma_l$ for the barrier penetration.   It is given explicitly by
\begin{eqnarray}
C(l,E)&=&-  \sum_{l'=0}^{l-1}
\frac{2l'+1}{1+\exp \{ (E -E_{l'})/\Gamma_{l'} \}}
\Theta(l-1)
\nonumber\\
&+& \sum_{l'=l+1}^{\infty} 
\frac{2l'+1}{1+\exp \{ (E_{l'} -E)/\Gamma_{l'} \}}.
\label{C7}
\end{eqnarray}
where $\Theta(x)=1$ for $x\ge 0$.

For the evaluation of the reaction cross section and the correction
term $C(l,E)$, we shall study a simple model in which we assume that the
barriers $E_l$ and the frequencies $\hbar \omega_l$ (or $\Gamma_l$)
are related to $l$ by
\begin{eqnarray}
E_l &=& E_0 +\frac{\hbar^2 l(l+1)}{2\mu  R_0^2},
\label{eq3}\\
\hbar \omega_l &\sim& \hbar \omega,~~ (~{\rm or}~~ \Gamma_l=\Gamma),
\label{eq4}
\end{eqnarray}
where $\mu=A_1 A_2 m_{\rm nucleon}/(A_1+A_2)$ is the reduced mass. We
shall further convert the summations in Eq. (\ref{C7}) as integrals in
the continuum approximation, then the correction term is given
explicitly by
\begin{eqnarray}
C(l=0,E)
&=&
\frac{2\mu R_0^2 \Gamma }{\hbar^2}\ln \biggl [ 
{1+\exp \{ (E -E_{1})/\Gamma \}}\biggr ],
\label{pc11}
\end{eqnarray}
\begin{eqnarray}
C(l=1,E)
&=&
\frac{2\mu R_0^2 \Gamma }{\hbar^2}\ln \biggl [ 
{1+\exp \{ (E -E_{2})/\Gamma \}}\biggr ]\nonumber\\
&-&  
\frac{1}{1+\exp \{ (E -E_{0})/\Gamma \}},
\label{pc12}\end{eqnarray}
\begin{eqnarray}
C(l\ge 2,E)
&=&
\frac{2\mu R_0^2 \Gamma }{\hbar^2} \ln \biggl [ 
{1+\exp \{ (E -E_{l+1})/\Gamma \}}\biggr ]\nonumber\\
&-&\frac{2\mu R_0^2 \Gamma }{\hbar^2}
\ln \biggl [ \frac{{1+\exp \{ (E_{l-1} -E)/\Gamma \}}}
{1+\exp \{ (E_{0} -E)/\Gamma \}}
\biggr ].~~~
\label{pc13}\end{eqnarray}
Thus, Eq.\ (\ref{pc7}), with supplementary equations (\ref{pc6}),
(\ref{pc11}), (\ref{pc12}), and (\ref{pc13}), gives the reaction cross
section as a function of energy $E$.

We can evaluate the reaction cross section $ ( {\sigma_r}/{\pi
  \slam^2} )(E_l)$ at the barrier $E_l$.  It is given by
\begin{eqnarray}
 ( {\sigma_r}/{\pi \slam^2} )\biggr |_{E_l} &=& l(l+1)+\frac{1}{2}+C(l,E_l),
\label{5} 
\end{eqnarray}
where the correction term $C(l,E_l)$ is 
\begin{eqnarray}
C(l=0,E_l)&=&
\frac{2 \Gamma}{E_1-E_0} \ln \biggl [ 
{1+\exp \{ (E_{l} -E_{l+1})/\Gamma \}}\biggr ], \nonumber\\
C(l=1,E_l)&=& 
\frac{2 \Gamma}{E_1-E_0}
\ln \biggl [ 
{1+\exp \{ (E_{l} -E_{l+1})/\Gamma \}}\biggr ]
\nonumber\\
&-& 
\frac{1}{1+\exp \{ (E_{l} -E_{0})/\Gamma \}}, \nonumber\\
C(l\ge 2,E_l)&=& 
\frac{2 \Gamma}{E_1-E_0}
 \ln \biggl [ 
{1+\exp \{ (E_{l} -E_{l+1})/\Gamma \}}\biggr ]\nonumber\\
&-&
\frac{2 \Gamma}{E_1-E_0}
\ln \biggl [ \frac{{1+\exp \{ (E_{l-1} -E_{l})/\Gamma \}}}
{1+\exp \{ (E_{0} -E_{l})/\Gamma \}}
\biggr ].~~~~
\label{eq10}
\end{eqnarray}
Equation (\ref{5}) has a simple physical interpretation.  As
illustrated in Fig.\ 2.1 of Blatt and Weisskopf \cite{Bla52}, the
partial wave $l'$ contributes $2l'+1$ units to the dimensionless
measure of the reaction cross section.  The total contribution is the
integral of $\int dl' (2l'+1)$.  Therefore, the dimensionless measure
of the reaction cross section, up to the fusion barrier of the $l$
partial wave, is given by $l(l+1)$ on the right hand side.  The
additional constant $1/2$ is purely quantum mechanical in origin and
it depends on the symmetry of the colliding system, as will be
discussed in Sec. VI.  The correction term $C(l,E_l)$ in
Eq.\ (\ref{5})  takes into account the finite energy
width $\Gamma_l$ for barrier penetration probability relative to the spacing
between adjacent fusion barriers.

Equations (\ref{5}) and (\ref{eq10}) can be inverted to provide the
rule for the ``barrier analysis" for unequal nuclei as follows.  The
fusion barrier $E_l$ for the $l$ partial wave is located at the value
of energy $E$ at which the dimensionless reaction cross section
measure, $\sigma_r/\pi\slam^2$, is equal to $l(l+1)$+1/2+$C(l,E_l)$. If
the dimensionless measure $({\sigma_r}/{\pi \slam^2})$ can be obtained
experimentally as a function of $E$, the heights of various fusion
barriers $E_l$ can be determined iteratively, within the present model
of fusion barrier penetration.

In the beginning of the iteration, one neglects the correction $C(l,E_l)$,
and the barriers $E_l$ can be determined from the $\sigma_r/\pi
\slam^2$ values.  With the knowledge of the barriers heights $E_l$ the
correction terms $C(l,E_l)$ can be evaluated for different $l$ partial
waves, and the barrier quantities $E_l$ can be corrected.  In these
iterations, it is necessary to know the width parameter $\Gamma$,
which can be obtained either from a fit of the experimental fusion
cross section with Eq.\ (\ref{eq1}), or from the penetration
probability analysis, as will be discussed in Sec. VII.

We list the values of $\sigma_r/\pi \slam^2$ at which the fusion
barriers $E_l$ are located for the collision of unequal nuclei in
Table I, when the correction term $C(l,E_l)$ can be neglected.

\begin{table}[h]
 \caption { The value of the dimensionless measure $\sigma_r/\pi
   {\protect\slam}^2$ at which the fusion barriers $E_l$ for the $l$
   partial wave is located, for the collision of unequal nuclei, 
when the correction term $C(l,E_l)$ can be neglected.
 }
\vspace*{0.2cm} 
\hspace*{0.0cm}
\begin{tabular}{|c|c|c|c|c|c|c|c|c|}
\hline    
  $l$   & 0 & 1 & 2 & 3 & 4 & 5   \\\hline    $\sigma_r/\pi\slam^2$ &~~ 0.5~~ & ~~2.5~~ &~~ 6.5~~ & ~~12.5~~ &~~ 20.5~~ & ~~30.5~~  \\
\hline
\end{tabular}
\end{table}

The ratio $\Gamma/(E_l-E_{l-1})$ in the correction term $C(l,E_l)$
varies with the colliding nuclei mass number as $A^{5/3}\Gamma/2l$.
Thus, a decrease in the mass number or an increase in the angular
momentum $l$ will lead to a smaller $\Gamma/(E_l-E_{l-1})$ and a
smaller correction term $C(l,E_l)$.  Our investigations in subsequent
sections (Tables II and III) indicate that for light nuclei
collisions, the condition of $\Gamma \ll |E_l-E_{l\pm 1}|$ is
approximately fulfilled so that $C(l,E_l)$ is small.  It becomes
appropriate to neglect the correction term in the barrier analysis for
light nuclei collisions.  For heavy nuclei collisions, while the
neglect of the correction $C(l,E_l)$ may be appropriate in the barrier
analysis for large angular momentum $l$, the correction term must be
properly taken into account for partial waves with small values of
$l$.

\section{ The Continuum Approximation and its Corrections }

Our model assumption of a linear dependence of $E_l$ on $l(l+1)$ in
Eq.\ (\ref{eq3}) is a reasonable concept for cases when the 
effective separation of the two colliding nuclei $R_0$ at the fusion
barrier is insensitive to the change of the angular momentum $l$.
While such an assumption is reasonable for most reactions, there are
however cases, such as ${}^{12}$C+${}^{13}$C, in which such an
assumption may not be valid.  The method of barrier analysis we
have just developed in Sec. II may be used to diagnose the
pathological case.  We shall discuss the collision of
${}^{12}$C+${}^{13}$C in Sec. V.

By replacing the sum in Eq. (\ref{eq2}) by an integral in the
continuum approximation, the reaction cross section can be integrated
to yield the analytical formula of Eq.\ (\ref{eq1}) \cite{Won73}.
Such a replacement of the discrete $l$ variable by a continuous
variable incurs errors.  It is desirable to find the magnitude of
the errors and ways to correct for these errors, if a higher accuracy
is required.  For brevity of notation, we introduce $a$ and $g$ to
rewrite $f_l$ as
\begin{eqnarray}
 f_l=f(l)&=& \frac{2l+1}{1+\exp \{ [a l(l+1) -\epsilon)]/\Gamma \}}
\equiv \frac{2l+1}{1+g } ,
\end{eqnarray}
where $g=\exp\{~[al(l+1)-\epsilon]~/~\Gamma\}$,
$\epsilon = E-E_0$,  and $a={\hbar^2}/{2\mu R_0^2}$.  

Our effort to examine the errors brings us to partition the
contributions in Eq.\ (\ref{3}) into two groups: (i) one group of $l$
states for which the continuum approximation is a reasonable concept
and analytical results can be readily obtained, and (ii) another group
of discrete $l$ states which remain as they are, without applying the
continuum approximation, and their contributions to the total reaction
cross section can be subsequently singled out for scrutiny.

The $l=0$ state is important in sub-barrier fusion and it is not
suitable for the continuum approximation.  We shall keep $f_0$ to
remain as it is in Eq.\ (\ref{3}).  We can express $f_l$ with $l\ge
1$ as a continuous integral with a correction $\Delta f_l$
\begin{eqnarray}
f_l 
&=& \int_{l-1/2}^{l+1/2} dl f(l) + \Delta f_l.
\label{eq8}
\end{eqnarray}
The function $f(l)$ has an indefinite integral
\begin{eqnarray}
 \int \!\!dl f(l)\!\! =\!\!\frac{-\Gamma}{a}
\ln \left \{ 1+\exp \left  [ \frac{\epsilon-a l(l+1)}{\Gamma}\right ]\right  \} \equiv \!\!F(l).
\end{eqnarray}
In terms of the function $F(l)$, we have
\begin{eqnarray}
f_l = F(l+1/2)-F(l-1/2) + \Delta f_l.
\label{eq9}
\end{eqnarray}
By definition, the correction term $\Delta f_l$ is then given by
\begin{eqnarray}
\Delta f_l = f(l)-[F(l+1/2)-F(l-1/2)].
\end{eqnarray}
Treating $l$ as a continuous variable in the above equation and
expanding the function $F(l\pm 1/2)$ about $l$ in a Taylor series with
$\Delta l=1/2$, we obtain explicitly
\begin{eqnarray}
\Delta f_l
&=&-2 \sum_{n=2,4,..}\frac{ (\Delta l)^{n+1} }{(n+1)! } \frac{d^{n} }{dl^{n}}f(l) .
\label{eq11}
\end{eqnarray} 
We thus obtain the central result that in the continuum
approximation, any term $f_l$ with $l \ge 1$ in the series of
Eq.\ (\ref{3}) can be replaced by Eq.\ ({\ref{eq9}) of the known
  function $F(l)$, with $\Delta f_l$ given by Eq.\ (\ref{eq11}).  For
  example, if we wish to partition the partial wave into those with
  $[0,l_L]$ as a discrete sum, with those in $[l_L+1,\infty]$ in the
  continuum approximation, then we obtain for such a partition
\begin{eqnarray}
\frac{\sigma_r}{\pi \slam^2}&=& \sum_{l=0}^{l_L}f(l)
-F(l_L+1/2) \nonumber\\
& &-2 \hspace*{-0.3cm}\sum_{l=l_L+1,l_L+2,..}\left \{ \sum_{n=2,4,..}\frac{ (\Delta l)^{n+1} }{(n+1)! }
 \frac{d^{n} }{dl^{n}}f(l) \right \}.~~~
\end{eqnarray}
We can use the relation
\begin{eqnarray}
\pi \slam^2 \frac{\Gamma}{a}=\pi R_0^2\frac{\Gamma}{E}
\end{eqnarray}•
to write the reaction cross section as 
\begin{eqnarray}
\sigma_r&
=& {\pi \slam^2 }{(f_0+f_1+...+f_L)}
\nonumber\\
&+ &
\pi R_0^2\frac{\Gamma}{E}\ln\left  \{ 1+ \exp \left [ \frac{\epsilon}{\Gamma}-\frac{(l_L+1/2)(l_L+3/2)a}{\Gamma} \right ] \right \} 
\nonumber\\
&-&
 2 {\pi \slam^2 }\sum_{l=l_L+1,l_L+2,..}\left \{ \sum_{n=2,4,..}\frac{ (\Delta l)^{n+1} }{(n+1)! }
 \frac{d^{n} }{dl^{n}}f(l) \right \} . 
\label{14}
\end{eqnarray}
In the special partition by singling out only the lowest $l_L$=0 wave
for special consideration, then up to the third order $(\Delta
l)^3$, we obtain
\begin{eqnarray}
\sigma_r&
=& \frac{\pi \slam^2 }{1+\exp\{\epsilon/\Gamma\}}
+\pi R_0^2\frac{\Gamma}{E}\ln\left  \{ 1+ \exp \left [ \frac{\epsilon}{\Gamma}-\frac{3a}{4\Gamma} \right ] \right \} 
\nonumber\\ 
& & - \frac{\pi \slam^2}{24} 
\sum_{l=1,2,3,..}
\frac{d^{2} }{dl^{2}}f(l),
\label{eq16}
\end{eqnarray}
where the derivative in the correction term is 
\begin{eqnarray}
\frac{d^2} {dl^2} f(l)
&=&
6(2l+1)\left ( \frac{a}{\Gamma}\right ) \left [ -\frac{g}{(1+g)^2}\right ]
\nonumber\\
& &+ (2l+1)^3\left ( \frac{a}{\Gamma}\right )^2 \left [-\frac{g}{(1+g)^2}
+\frac{2g^2}{(1+g)^3}\right ].~~~
\end{eqnarray}
Terms on the right-hand side of Eq.\ (\ref{eq16}) have direct physical
meanings.  The first term corresponds to the contribution from the
lowest $l=0$ partial wave, and the second term corresponds to the
contribution from $l\ge 1$ partial waves in the continuum
approximation, and the last term is the correction due to the
continuum approximation up to the third order in $\Delta l$=1/2.

The above considerations can be generalized.  For the most general case \cite{Nef07},
\begin{eqnarray}
\frac{\sigma_r}{\pi \slam^2}
&=& \sum_{l_\nu}
[1+\eta (-1)^{l_\nu}]  f(l_\nu),
\end{eqnarray}
where (i) $\eta=0$ and $l_\nu=\nu=0,1,2,3,...$ for unequal nuclei,
(ii) $\eta=1$ and $l_\nu=2\nu=0,2,4,...$ for identical spin-0 nuclei
or for identical spin-1/2 nuclei with symmetric spatial and
antisymmetric spin wave functions, and (iii) $\eta=-1$ and
$l_\nu=2\nu+1=1,3,5,..$ for identical spin-1/2 nuclei with
antisymmetric spatial and symmetric spin wave functions.  The sum over
$l_\nu$ can be converted into a sum over $\nu$ with $\nu=0,1,2,...$ We
obtain up to the third order $(\Delta \nu)^3$ with $\Delta \nu=1/2$,
\begin{eqnarray}
\sigma_r
&=&
 \pi \slam^2 \frac{dl_\nu}{d\nu}
\frac{2l_0+1}{1+\exp \{ [a l_0(l_0+1) -\epsilon)]/\Gamma \}}
\nonumber\\
& &+\pi R_0^2\frac{\Gamma}{E}\ln\left  \{ 1+ \exp \left [ \frac{\epsilon}{\Gamma}- \frac{l_{1/2}(l_{1/2}+1)a}{\Gamma} \right ] \right \}
\nonumber\\
& &  
-2\pi \slam^2\left ( \frac{dl_\nu}{d\nu} \right)^3  \left (\sum_{\nu=1,2,3,...}\frac{ (\Delta \nu)^3 }{3! } \frac{d^2}{dl_\nu^2} f(l_\nu) \right )\!.~~~~~
\label{eq18}
\end{eqnarray}

\section{Numerical Results and Comparison with Data}

In presenting our numerical results, we shall label Eq.\ (\ref{eq1})
as Formula I, the sum of the first two terms in Eq.\ (\ref{eq16}) [or
  (\ref{eq18})] as Formula II, and the sum of all three terms in
Eq.\ (\ref{eq16}) [or (\ref{eq18})] as Formula III.  In simple
physical terms, Formula I corresponds to the earlier result of
Ref.\ \cite{Won73} using the continuum approximation for all
partial waves.  Formula II is obtained by writing out the contribution
from the lowest $l=0$ partial wave explicitly and treating the higher $l\ge 1$
partial wave contributions  in the continuum
approximation. Formula III involves Formula II with the
inclusion of corrections up to the third order in $\Delta l = 1/2$.
Following Esbensen \cite{Esb12}, we shall label the cross section
obtained in the sum of Eq.\ ({\ref{eq2}) over the Hill-Wheeler
  penetration probability, under the assumption of Eqs.\ (\ref{eq3}) and
  (\ref{eq4}), as the Hill-Wheeler cross section.

\begin{figure} [h]
\includegraphics[scale=0.45]{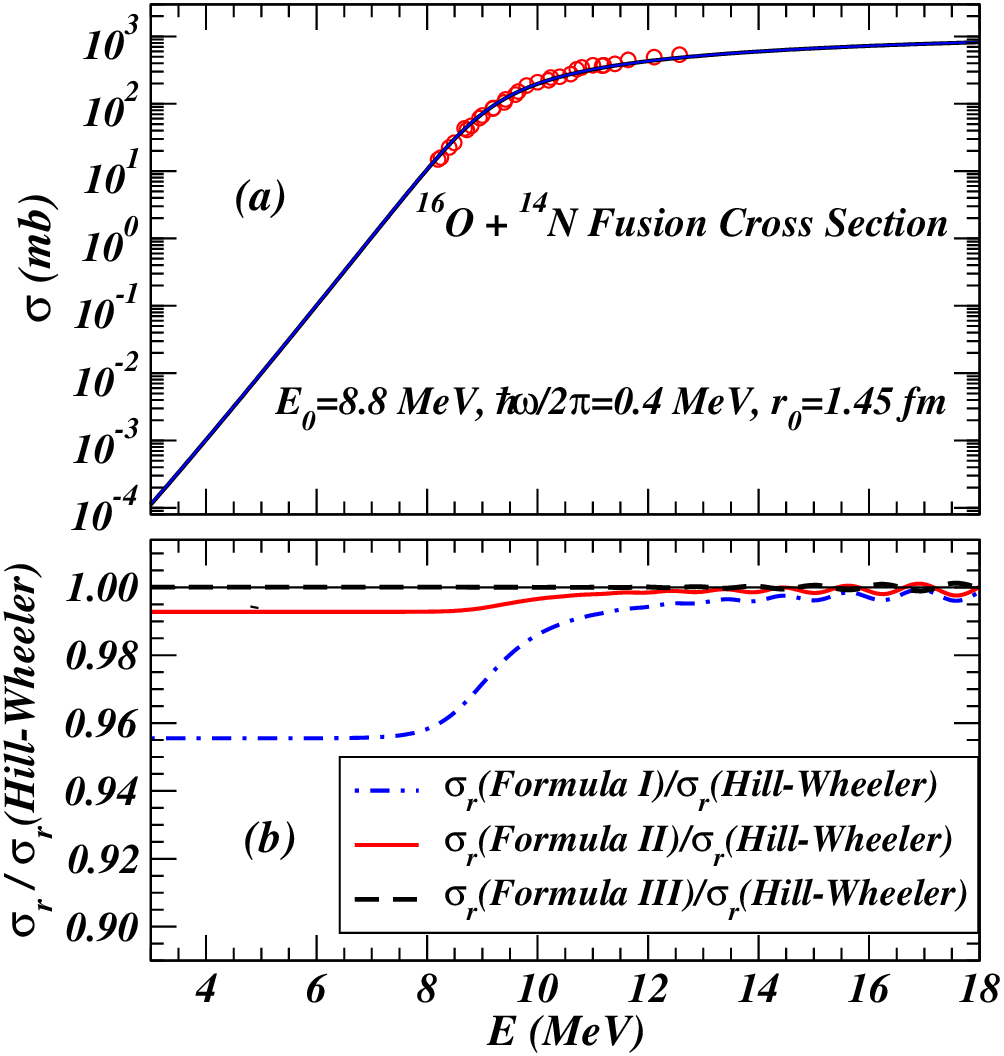}
\caption{(Color online) (a) The comparison of experimental fusion
  cross section for $^{16}$O + ${}^{14}$N \cite{Swi77} with
  theoretical results. (b) The ratio of the cross sections from
  Formulas I, II, and III relative to the Hill-Wheeler cross section. }
\end{figure}

We examine the sample case for the collision of ${}^{16}$O+${}^{14}$N
where the experimental data \cite{Swi77} are shown in Fig.\ 1(a).  We
show the fit to the fusion cross section obtained with $E_0=8.8$ MeV,
$\Gamma=0.40$ MeV ($\hbar\omega=2.51$ MeV), and $r_0=1.45$ fm as
curves in Fig.\ 1(a).  The differences among the three formulas
cannot be distinguished in the logarithmic plot.  In order to see the
differences, we plot the corresponding ratios of the cross sections
relative to the Hill-Wheeler cross section in Fig.\ 1(b).  In the
sub-barrier region, we find that Formula I gives an error of about
4.5\%, Formula II gives an error of less than 1\%, and Formula III gives
an error of less than 0.01 \%.  In the high energy region, all three
formulas give small errors, of the order of at most 0.3\%.

We conclude from the results of Fig.\ 1 that for unequal nuclei
Formula I is adequate for the sub-barrier region if errors of 5\% are
permitted, Formula II gives a more accurate result in all regions with
less than 1\% error, and Formula III gives even smaller errors in all
regions.

\begin{figure} [h]
\includegraphics[scale=0.45]{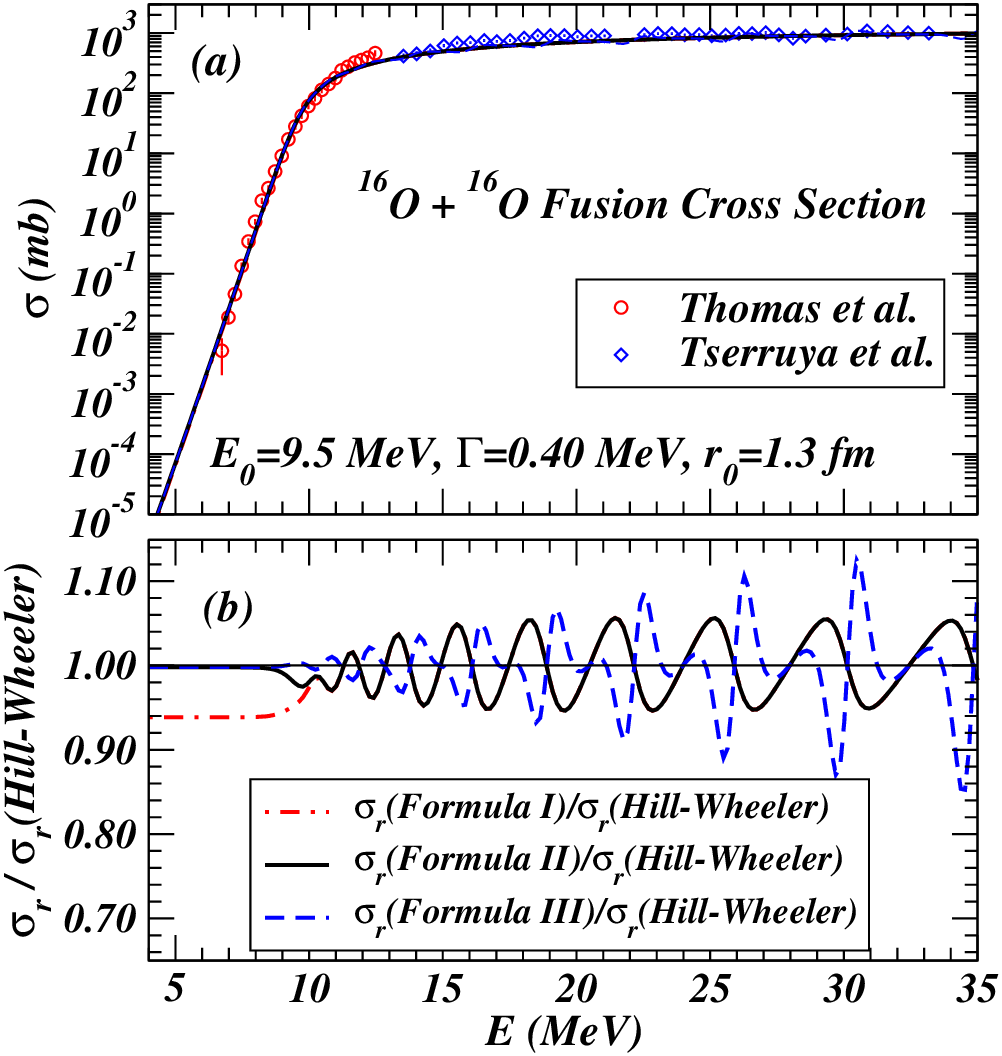}
\caption{(Color online) (a) Comparison of fusion cross section data
  for the collision of identical spin-0 nuclei $^{16}$O +${} ^{16}$O
  \cite{Tho86,Tse78} with theoretical curves.  (b) The ratio of cross
  sections from Formulas I, II, and III relative to the
  Hill-Wheeler cross section.}
\end{figure}

 In Fig.\ 2, we examine next the fusion cross section for the
 collision of ${}^{16}$O+${}^{16}$O where the experimental data
 \cite{Tho86,Tse78} are shown in Fig.\ 2(a) and the theoretical
 results from the three different formulas calculated with the
 parameters $E_0$=9.5 MeV, $\Gamma$= 0.4 MeV, and $r_0$=1.3 fm are shown
 as curves.  In this case with identical spin-0 nuclei with spatially
 symmetric wave functions, only the even-$l$ partial waves contribute
 to the reaction cross section.  Formula II consists of the first two
 terms on the right-hand side of Eq.\ (\ref{eq18}) and Formula III
 consists of all three terms in Eq.\ (\ref{eq18}).

On the logarithmic scale of Fig.\ 2(a), the results from all three
formulas cannot be well distinguished.  The agreement of the
experimental data with the theoretical curves may appear reasonable.
However, in Fig. 2(b) we examine the ratio of the cross section
obtained with the three different formulas relative to the
Hill-Wheeler cross section.  In the sub-barrier region, Formula I
gives errors of order 6\%, Formulas II and III give errors of less than
0.3\%.

In the high-energy region, all three formulas
 give errors oscillating regularly about zero as a function of $E$.
 The magnitude of the oscillation is nearly constant for Formula II at
 high energies, but it increases as the energy increases for Formula III.
 These results are in agreement with the earlier observation of
 Esbensen \cite{Esb12}, who noted that, as a result of the spatial symmetry of the
 wave function such that only even $l$ states contribute, the energy
 separation between the contributing $l$ state and the $l+2$ state increases
 as energy increases, and the total reaction cross section exhibits a
 step-wise increase when a high-$l$ state enters into the formation of
 a fused system.  As a consequence, the continuum approximation  contains
 large and oscillating errors.  In mathematical terms, the large error
 arises from the fact that even though the expansion parameter $\Delta
 \nu=1/2$ is less than unity in Eq.\ (\ref{eq18}), it is multiplied by
 the factor $dl_\nu/d\nu$ with $dl_\nu/d\nu=2$.  Thus the
 effective expansion parameter is $(\Delta \nu)(dl_\nu/d\nu) =1$ and the
 expansion in Formula III does not properly converge.

We conclude from Fig. 2 that, for the collision of identical spin-0
nuclei at high energies, the continuum approximation incurs large
errors.  Formulas I and II give errors of about 5\% while Formula III
gives even greater errors up to 15\%.  On the other hand, near the
sub-barrier region Formula II gives very small errors.

\section{Barrier Analysis for  ${}^{12}$C+${}^{13}$C }}

The results in the last few sections pertain to the collisions of
both light and heavy nuclei.  In the collision of heavy nuclei,
however, there is the complication that the correction term $C(l,E_l)$
for low-$l$ partial waves for the barrier analysis must be properly
taken into account in an iterative procedure, as specified by
Eqs.\ (\ref{5}) and (\ref{eq10}).  In contrast, for light nuclei
collisions, the width parameter $\Gamma$ is found to be substantially
smaller than the separation between adjacent barriers so that these
correction terms can be neglected in the barrier analysis, leading to
a great simplification of the problem.  For  simplicity, we
shall therefore specialize to light nuclei collisions in subsequent
sections.

Our ability to reach the simple results in the last sections relies on
the assumption that the fusion barrier $E_l$ for the $l$
partial wave is a linear function of $l(l+1)$, as given by
Eq.\ (\ref{eq3}).  There may be nuclear collisions in which such an
assumption may not be valid.

\begin{figure} [h]
\includegraphics[scale=0.48]{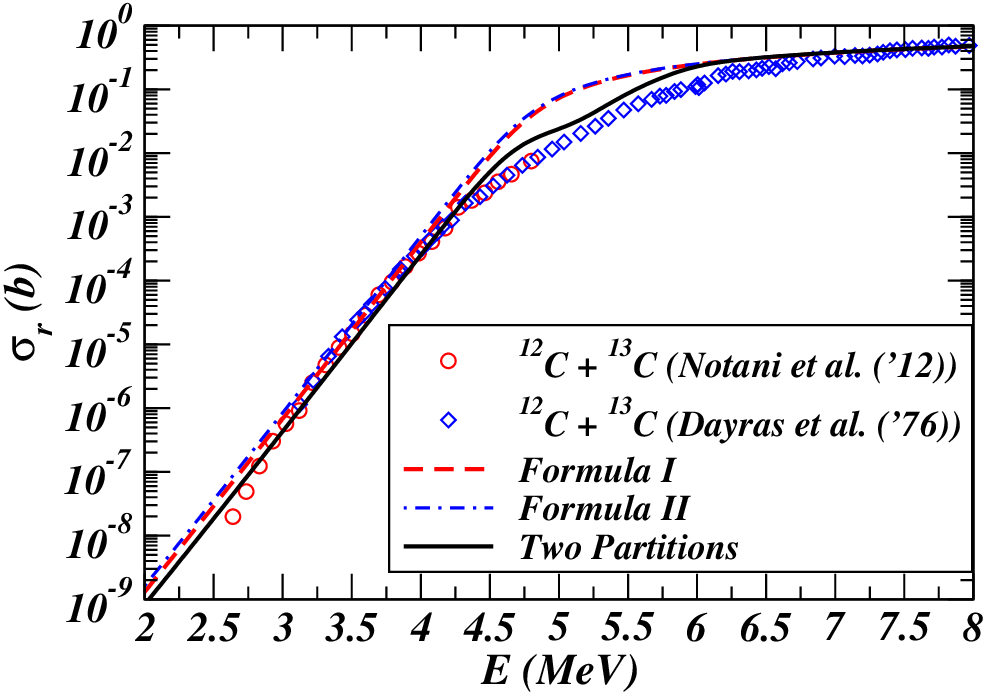}
\caption{(Color online) Comparison of fusion cross section data
  for $^{12}$C + ${}^{13}$C \cite{Not12,Day76} with theoretical
  curves.  }
\end{figure}

We examine ${}^{12}$C+${}^{13}$C where the data are shown in Fig. 3.
The data can be explained well by a coupled-channels calculation with
the ingoing wave boundary condition (IWBC) and the M3Y+repulsive
potential \cite{Not12}.  Nevertheless, it is useful to examine these
data from a complementary perspective in barrier-penetration points of
view.  One then finds that the ${}^{12}$C+${}^{13}$C data
\cite{Not12,Day76} cannot be described by Formula I, II or III.  Any
fit to the data near the threshold will miss the data at some other
energy region.  We show in Fig.\ 3 the results of Formula I and II
obtained with the parameters $E_0=4.7$ MeV, $\Gamma=0.15$ MeV, and
$r_0$=1.3 fm as the dashed curve and the dash-dotted curve, respectively.
The region around $E\sim$ 5 MeV is not well reproduced.  From the
viewpoint of barrier penetration and the simple model with the
assumption of $E_l = E_0 + a l(l+1)$ in Eq.\ (\ref{eq3}), the shape of
the fusion cross section in the collision of ${}^{12}$C+${}^{13}$C
poses a problem.

\begin{figure} [h]
\includegraphics[scale=0.45]{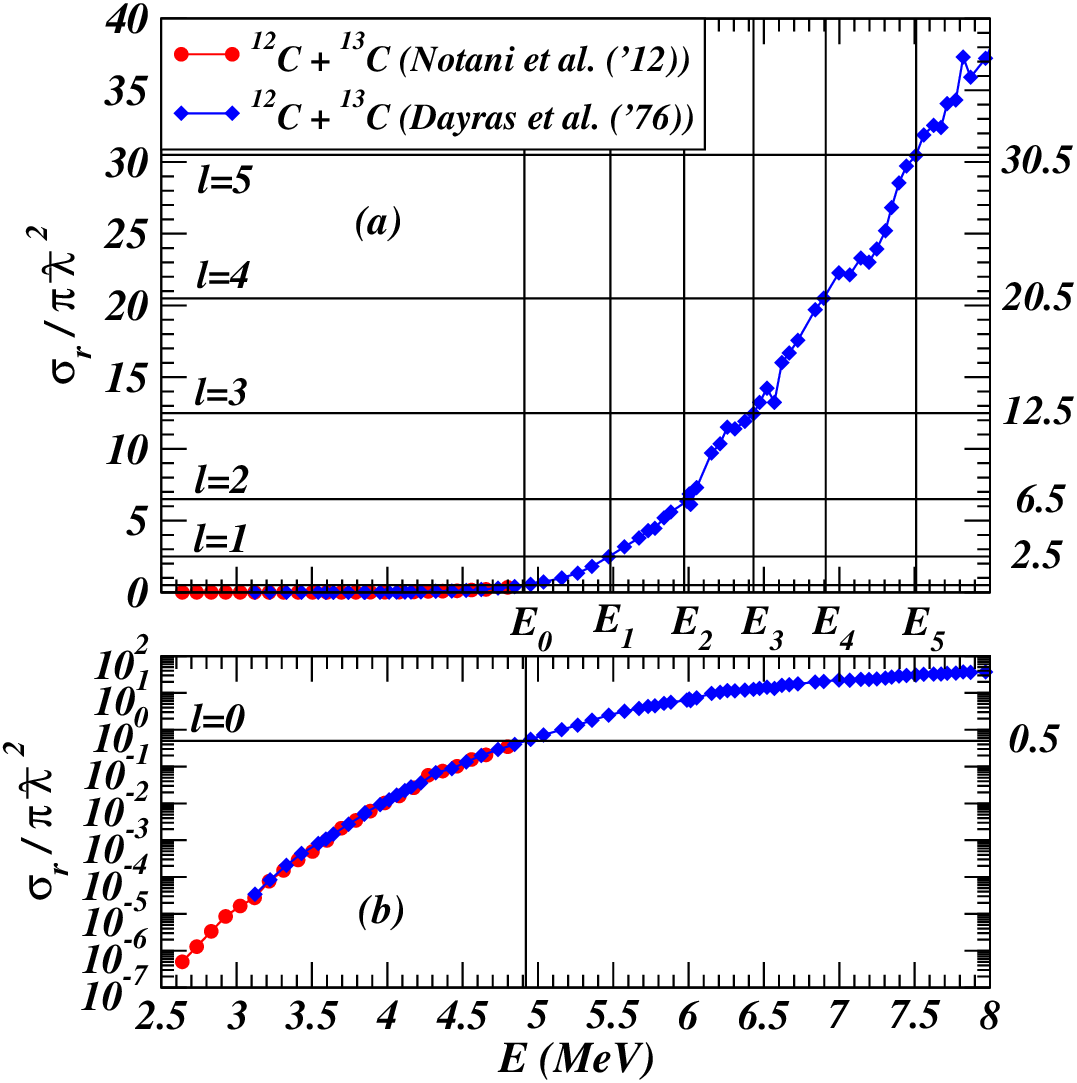}
\caption{(Color online) The dimensionless quantity $(\sigma_r / \pi
  {\protect\slam}^2)$ as a function of $E$, on a linear scale (a), and on a logarithmic scale (b) for the
  collision of ${}^{12}$C+${}^{13}$C.  In the barrier analysis,
  the value of $E$ at which $(\sigma_r/ \pi {\protect \slam}^2)$ is
  $l(l+1)+1/2$ is the fusion barrier $E_l$.  The horizontal lines
  indicate the $l(l+1)+1/2$ values shown as numbers given along the
  right vertical axis.  The vertical lines indicate the positions of
  the fusion barriers $E_l$ at which $\sigma_r/\pi {\protect
    \slam}^2=l(l+1)+1/2$.  Data points are from \cite{Not12,Day76}.  }
\end{figure}

To check whether the assumption Eq.\ (\ref{eq3}) is valid for
${}^{12}$C+${}^{13}$C, we can carry out a ``barrier analysis"
by plotting $\sigma_r/\pi \slam^2$ as a function of $E$ as shown in  Fig. 4.
The plots in  Fig.\ 4(a) are  on a linear scale and those of Fig.\ 4(b) on a logarithm scale.
The rule in Eq.\ (\ref{5}) stipulates that the barrier $E_l$ is
the value of energy $E$ at which $\sigma_r/\pi\slam^2$ is
$l(l+1)+1/2+C(l,E_l)$.  For light nuclei collision
for which the width $\Gamma$ is substantially 
smaller than the spacing between adjacent
barriers, the correction term $C(l,E_l)$ is small and can be neglected. 
We plot $\sigma_r/\pi\slam^2=l(l+1)+1/2$ as horizontal
lines in Fig.\ 4.  The energy values $E$ where the horizontal lines
meet the data points give the locations of the fusion barriers $E_l$
in Fig.\ 4.

 On plotting the barrier $E_l$ obtained in the analysis of
 Fig.\ 4 as a function of $l(l+1)$, one observes in Fig.\ 5 that $E_l$
 is not a linear function of $l(l+1)$, as assumed in Eq.\ (\ref{eq3}).
 While the linear relationship is reasonable for $l\agt 2$, the
 systematics of $E_l$ for $l\le 2$ appears to be different from those
 with $l \agt 2$.
\begin{figure} [h]
\includegraphics[scale=0.45]{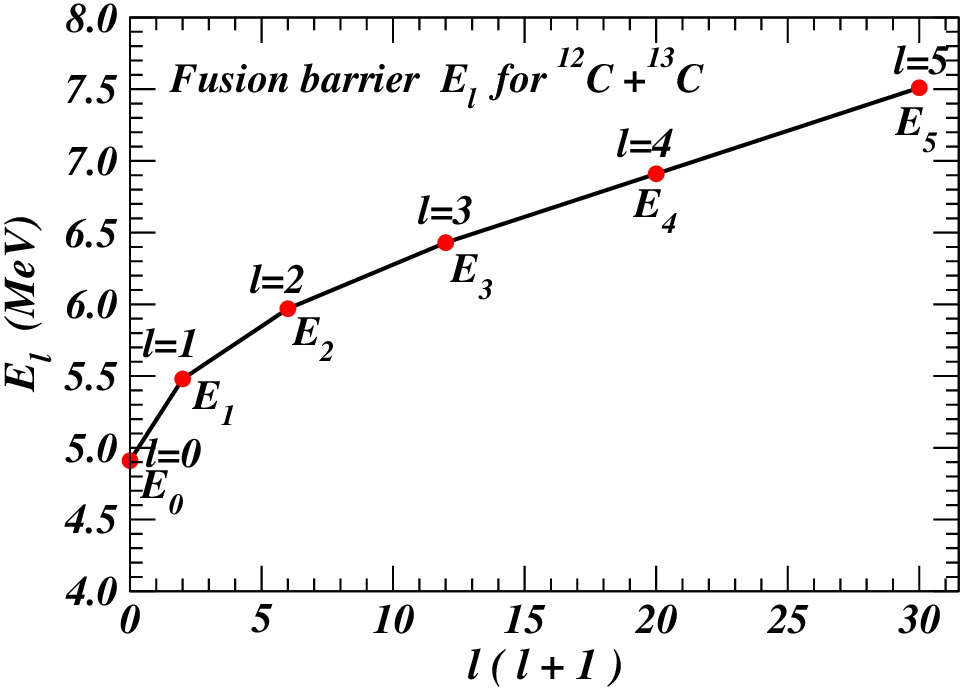}
\caption{(Color online) The fusion barrier $E_l$ 
as a function of $l(l+1)$
in the collision of  ${}^{12}$C+${}^{13}$C. }
\end{figure}

To examine the problem of barrier penetration, we partition the
partial waves into two parts in the intervals $[0,l_L]$ and
$[l_L$+1,$\infty]$.  We assume the Hill-Wheeler penetration
probability for the partition in $[0,l_L]$, and describe the cross
section from the partition $[l_L$+1,$\infty]$ by the continuum
approximation.  As given by Eq.\ (\ref{14}) the reaction cross section
with the neglect of the correction term and the assumption of
$\Gamma_l=\Gamma$ is
\begin{eqnarray}
\sigma_r
= {\pi \slam^2 }\sum_{l=0,1,..,l_L} \frac{2l+1}{1+\exp\{(E_l-E)/\Gamma\}}\hspace*{1.2cm}
\nonumber\\
\hspace*{-0.4cm}+
\pi R_0^2\frac{\Gamma}{E}\ln\left  \{ 1+ \exp \left [ \frac{\epsilon}{\Gamma}-\frac{(l_L+1/2)(l_L+3/2)a}{\Gamma} \right ] \right \}. 
\label{22}
\end{eqnarray}

Figure 5 indicates that the dependence of the fusion barriers $E_l$ on
$l(l+1)$ deviate from a linear relationship, for the first three
partial waves with $l=0,1$ and 2.  Consequently, we partition the
partial waves into two partitions of $[0,2]$ and $[3,\infty]$ with
$l_L=2$.  The values of $E_0,E_1,$ and $E_2$ can be read off from
Fig.\ 5 as the starting point for parameter search, with minor fine
tuning.  The results with $E_0=4.7$ MeV, $E_1=5.4$ MeV, $E_2=5.85$
MeV, $\Gamma=0.15$ MeV, and $r_0=1.3$ fm are shown as the solid curve
in Fig. 3.  We observe that although the fit of Eq.\ (\ref{22}) to the
experimental data is not perfect, the agreement with experimental data
around $E\sim5$ MeV is substantially improved.  The simple comparison
indicates that a possible solution of the peculiar shape of the fusion
cross section may involve fusion barriers increasing in a
non-linear way as a function of $l(l+1)$, corresponding to a fusion
radial distance occurring at a much reduced separation for the lowest
partial waves in ${}^{12}$C+${}^{13}$C collisions.  This may be
related to the need for a repulsive core in the interaction and the
mutual excitation between the colliding nuclei as shown in
\cite{Not12}.

\section{Barrier Analysis for the Collision of Identical Light Nuclei}

The results in the last section illustrate the application of the rule
for barrier analysis in the collision of unequal nuclei.  The
reaction cross section for the collision of identical nuclei will need
to obey the symmetry of the total wave function with respect to the
interchange of the colliding nuclei.  As a consequence, the barrier
height analysis rule will be modified for the collisions of identical
nuclei.

For the collision of identical spin-0 nuclei, the quantity 
$(\sigma_r/\pi \slam^2) $ is given by
\begin{eqnarray}
(\sigma_r/\pi \slam^2)=2 \sum_{l=0,2,4,...}
   \frac{2l+1}{1+\exp \{ (E_l -E)]/\Gamma_l \}}.
\end{eqnarray}
For light nuclei collisions,  $|E_l-E_{l\pm 1}| \gg \Gamma_l$ and
we can evaluate the
above quantity $(\sigma_r/\pi \slam^2)$ at the fusion barrier $E_l$
analytically.  For the collision of identical spin-0 nuclei, the
quantity $(\sigma_r/\pi \slam^2)$ at the fusion barrier $E_l$ (with
even $l$ value) is
\begin{eqnarray}
(\sigma_r/\pi \slam^2)\biggr |_{E_{l}}&=&l(l+1)+1.
\end{eqnarray}
For the collision of identical spin-1/2 nuclei, the total spin can be
$S=0$ or $S=1$, with a weight of $1/4$ and $3/4$ respectively.  As a
consequence, the reaction cross section is given by
\begin{eqnarray}
(\sigma_r/\pi \slam^2)&=& 2\times \frac{1}{4} \sum_{l=0,2,4,...}\frac{2l+1}{1+\exp \{ (E_l -E)]/\Gamma_l \}} 
\nonumber\\
&+&2\times \frac{3}{4}
\sum_{l=1,3,5,7,...}  \frac{2l+1}{1+\exp \{ (E_l -E)]/\Gamma_l \}}.~~~~~~~
\end{eqnarray}
The barrier analysis rule is different for $E_l$ with even-$l$
or odd-$l$ partial waves.  Assuming $|E_l-E_j| \gg \Gamma_l$ for $l\ne
j$ for light nuclei collisions, we find that for the collision of identical spin-1/2 nuclei, the
quantity $(\sigma_r/\pi\slam^2)$ at the barrier $E_l$ with even $l$ is
given by
\begin{eqnarray}
(\sigma_r/\pi \slam^2)\biggr |_{E_{l}}
=l(l+1)+\frac{1}{4},
\end{eqnarray} 
and the quantity
$\sigma_r/\pi\slam^2$ at the barrier $E_l$ with odd 
$l$ is given by 
\begin{eqnarray}
(\sigma_r/\pi \slam^2)\biggr |_{E_{l}}
=l(l+1)+\frac{3}{4}.
\end{eqnarray}
These equations can be utilized to determine the fusion barriers from
experimental $(\sigma_r/\pi \slam^2)$ data for the collision of
identical or non-identical light nuclei.

We can summarize the rules for the barrier analysis as follows.
The dimensionless measure of the reaction cross section $(\sigma_r/\pi
\slam^2)$ at the fusion barrier $E_l$ is
\begin{eqnarray}
(\sigma_r/\pi \slam^2)\biggr |_{E_{l}}
=l(l+1)+K,
\end{eqnarray}
where $K$ is given by
\begin{eqnarray}
K=
\begin{cases}
\frac{1}{2}  & {\rm non-identical ~nuclei}, \cr
1                   & {\rm identical~spin ~0 ~nuclei}, \cr
\frac{1}{4}  & {\rm even-}l, {\rm~identical~spin~} \frac{1}{2}{\rm~nuclei}, \cr
\frac{3}{4}  &{\rm odd-}l, {\rm~identical~spin~ }\frac{1}{2}{\rm~ nuclei}.  \cr
\end{cases}
\end{eqnarray}
The differences of $(\sigma_r/\pi \slam^2)|_{E_l}$ in the different
cases are large for the lowest $l=0$ partial wave.  The differences of
$(\sigma_r/\pi \slam^2)|_{E_l}$ in the different cases are small, in
comparison with the first term $l(l+1)$, when $l$ is large.

\section{Penetration Probability Analysis and Resonances for Light Nuclei Collisions}

The penetration probability $P_l(E)$ for the $l$ partial wave is a
physical quantity that reveals important information on the dynamics
of the fusion process.  The energy $E$ at which $P_l(E)$ is $1/2$ is
at the top of the fusion barrier, and the shape of the potential
barrier is governed by the shape and the energy dependence of
$P_l(E)$.  It is desirable to extract such a quantity from
experimental data for light nuclei collisions for which the width 
$\Gamma$ is substantially smaller than the separation between adjacent 
barriers. 

\subsection{Collision of unequal light nuclei}

We shall consider first the collision of unequal light nuclei and express
the dimensionless cross section $(\sigma_r/\pi \slam^2)$ in terms of
the penetration probability $P_l(E)$ as
\begin{eqnarray}
\frac{\sigma_r}{\pi \slam^2}
&=& \sum_{l=0,1,2,3,..}  (2l+1) P_l(E)
\label{pen}
\end{eqnarray}
The dimensionless cross section $(\sigma_r/\pi \slam^2)$ appears so
frequently that it is appropriate to abbreviate it by $\Sigma(E)$ that
is explicitly a function of the energy $E$.

The penetration probability $P_l(E)$ can be extracted from the
dimensionless cross section $\Sigma(E)=(\sigma_r/\pi \slam^2)$ if we
assume that the contributions of different partial waves to the
dimensionless cross section are well separated in energy as in light nuclei collisions.  Under such
an assumption, we can consider the contributions to $P_l(E)$ from
different partial waves.  In the domain of $E$ in which $P_l(E)$ is
significant, the contribution from each of the lower $l'< l$ partial
waves is saturated to $P_{l'}(E)=1$ while the contribution form each
of the higher $l'>l$ partial waves is negligible.  We can 
decompose the sum over $l$ in Eq.\ (\ref{pen}) into individual
contributions.  For the $l$ partial wave in the collision of unequal
nuclei, the penetration probability is then given by
\begin{eqnarray}
P_l(E) = \frac{\Sigma(E)-B(l)}{2l+1} \Theta[T(l)-\Sigma(E)]\Theta[\Sigma(E)\!-\!B(l)],\nonumber
\label{aa}\\
\end{eqnarray}
where $\Theta$ is the step function, $ T(l)$ is the top delimiter
of $\Sigma(E)$, and $B(l)$ is the bottom delimiter of $\Sigma(E)$.
For unequal nuclei collisions, the sum of $l$ is over $l=0,1,2,3,..$,
and the delimiters can be shown to be
\begin{eqnarray}
T(l)&=&(l+1)^2,\\
 B(0)&=&0 ~~~ {\rm and~~} B(l)=T(l-1)~~{\rm for~~} l\ge 1  .
\end{eqnarray}
If $\Sigma(E)=(\sigma_r/\pi \slam^2)$ is measured experimentally as a
function of $E$, the qpenetration probability $P_l(E)$ for different
partial waves can be determined.
\begin{figure} [h]
\includegraphics[scale=0.45]{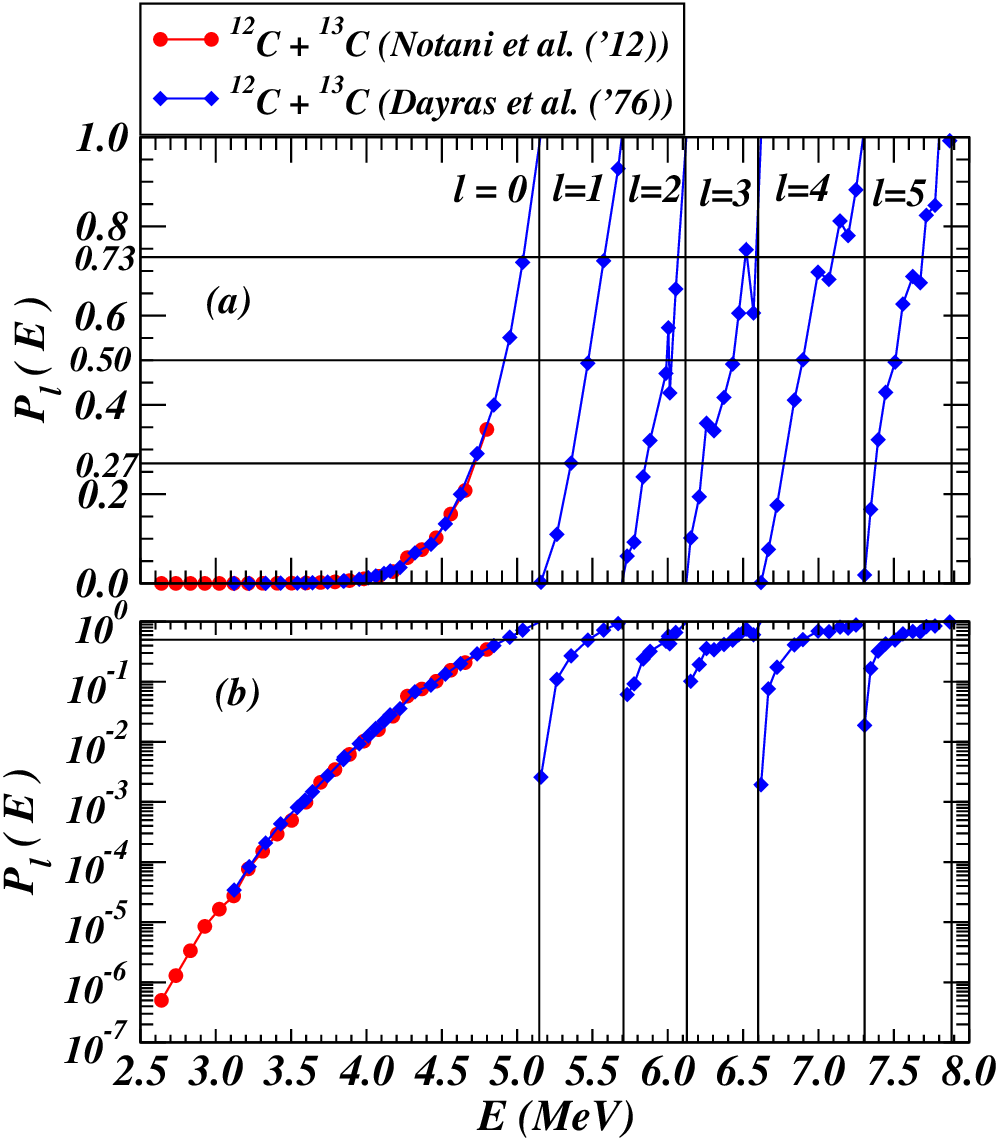}
\caption{(Color online) The penetration probability $P_l(E)$ as a
  function of $E$ on a linear scale (a) and on a logarithmic
  scale (b) for different partial waves $l$ extracted from
  the data of \cite{Not12,Day76}, for the collision of ${}^{12}$C+
  ${}^{13}$C.  }
\end{figure}

In Fig.\ 6 we show the penetration probability $P_l(E)$ as a function
of $E$ for various $l$ partial waves in the collision of ${}^{12}$C+
${}^{13}$C, obtained by using Eq.\ (\ref{aa}) and data from
\cite{Not12,Day76}.  For a given $l$, it is possible to determine the
fusion barrier $E_l$ as the energy at which $P_l(E)=0.5$, as
discussed in an equivalent procedure in Sec.s II and V.  We can
also extract an empirical width $\Gamma_l$ where $2\Gamma_l$ is
defined as the separation of $E$ between $P_l(E)=1/(1+e^{-1})=0.731$
and $P_l(E)=1/(1+e)=0.269$.  This empirical $\Gamma_l$ would be the
same as the $\Gamma_l$ in the Hill-Wheeler formula, if the penetration
probability follows the Hill-Wheeler formula.

In Table II, we list the fusion barrier $E_l$ and the width $\Gamma_l$
extracted from $P_l(E)$ in such a procedure for ${}^{12}$C+
${}^{13}$C.  As one observes, the widths for most of the partial waves
are about equal to $0.15$ MeV except for the $l=1$ and 2 partial waves,
which are about 0.11-0.12 MeV.

\begin{table}[h]
 \caption { The empirical values of $E_l$ and $\Gamma_l$ from
   $P_l(E)$, as extracted from the data of \cite{Bec81} for the
   collision of ${}^{12}$C+ ${}^{13}$C.  }
\vspace*{0.2cm} 
\hspace*{0.0cm}
\begin{tabular}{|c|c|c|c|c|c|c|c|c|}
\hline    
  $l$   & 0 & 1 & 2 & 3 & 4 & 5   \\\hline    
$E_l$ (MeV)  &~ 4.91~ & ~5.48~ &~ 5.97~ & ~6.43~ &~6.91~ & ~7.51~  \\
\hline
$\Gamma_l$ (MeV)  &~ 0.16~~ & ~0.11~~ &~~ 0.12~~ & $\sim$0.16~ &$\sim$0.15~~ & $\sim$0.16~ \\
\hline
\end{tabular}
\end{table}

\subsection{Collision of Identical Spin-0  Nuclei}

\begin{figure} [h]
\includegraphics[scale=0.45]{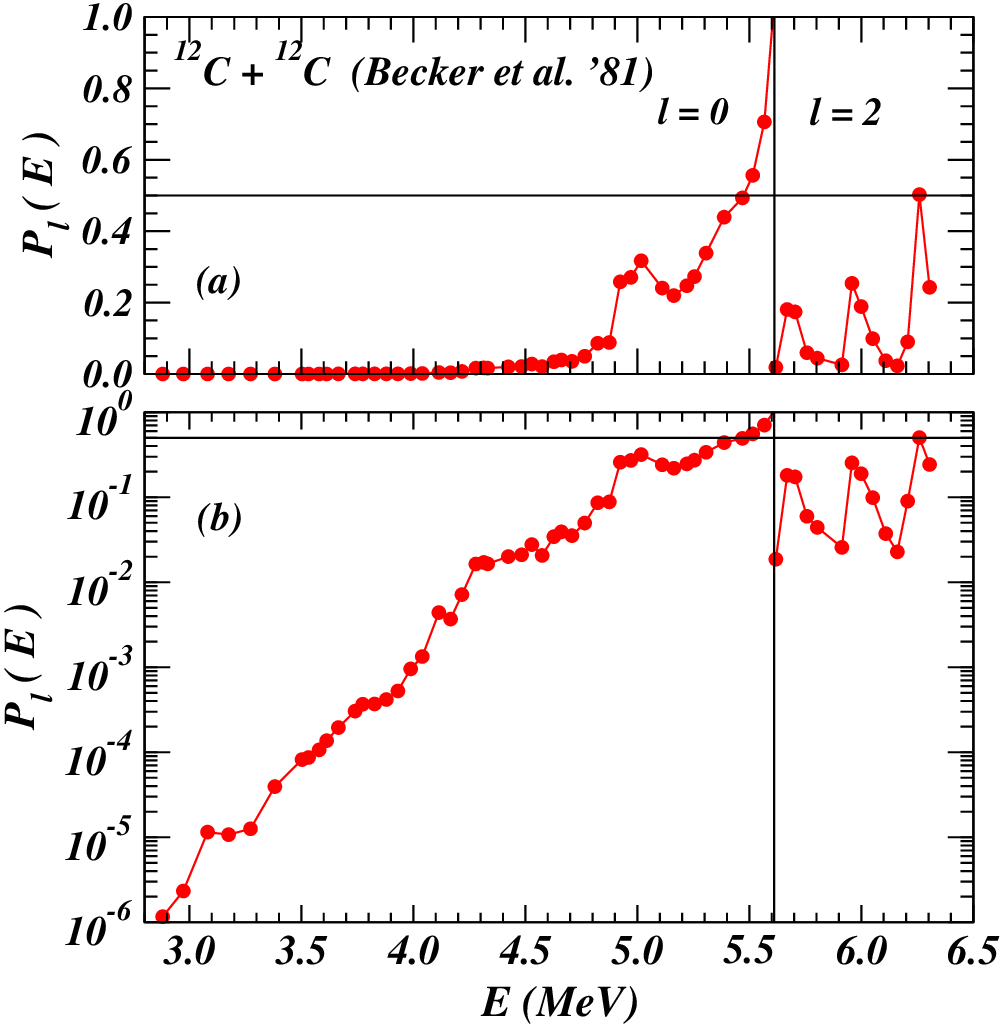}
\caption{(Color online) The penetration probability $P_l(E)$ as a
  function of $E$ on a linear scale (a) and on a logarithmic scale (b) extracted from the ${}^{12}$C+ ${}^{12}$C data of
  \cite{Bec81}.  }
\end{figure}

We shall consider next the collision of two identical spin-0
nuclei. The dimensionless cross section $(\sigma_r/\pi \slam^2)$ in
terms of the penetration probability $P_l(E)$ is
\begin{eqnarray}
\frac{\sigma_r}{\pi \slam^2}
&=& 2 \sum_{l=0,2,4,6,..}  (2l+1) P_l(E).
\end{eqnarray}
Under the assumption that $|E_l-E_j| \gg \Gamma_l$ for $l\ne j$
 for light nuclei collisions, the
contributions of different partial waves to the dimensionless cross
section are well separated in energy.  For the even-$l$ partial wave
in the collision of equal spin-0 light nuclei, the penetration probability is
given by
\begin{eqnarray}
P_l(E) = \frac{\Sigma(E)-B(l)}{2(2l+1)} \Theta[T(l)-\Sigma(E)]\Theta[\Sigma(E)-B(l)],\nonumber
\\\label{bb}
\end{eqnarray}
where we find
\begin{eqnarray}
T(l)&=&l(l+1)+2l+2,\\
B(0)&=&0,   ~~~ {\rm and~~}   B(l)=T(l-2)~~{\rm for~~} l\ge 2  .
\end{eqnarray}

The penetration probability $P_l(E)$ extracted from the experimental
${}^{12}$C+${}^{12}$C data \cite{Bec81} and Eq.\ (\ref{bb}) is shown
on a linear scale in Fig.\ 7(a), and on a logarithmic scale in
Fig.\ 7(b).  One finds the fusion barrier for the $l=0$ partial wave
$E_0$ at 5.46 MeV at which $P_{l=0}(E)=0.5$.  The boundary between the
$l=0$ and $l=2$ partial waves is approximately at $E$=5.6 MeV.  One
observes that $P_l(E)$ exhibits resonances.  The
resonances below $E=5.6$ MeV are most likely $l=0$ resonances whereas
those resonances above $E=5.6$ MeV are most likely $l=2$ resonances.

\subsection{Collision of identical spin-1/2  nuclei}

We shall consider next the collision of two identical spin-1/2
nuclei. The dimensionless cross section $(\sigma_r/\pi \slam^2)$
written in terms of the penetration probability $P_l(E)$ is
\begin{eqnarray}
\frac{\sigma_r}{\pi \slam^2}
&=& 2\left [ \frac{1}{4} \sum_{l=0,2,4,6,..}  
+ \frac{3}{4}\sum_{l=1,3,5,7,..}\right ]
(2l+1) P_l(E).~~~~~~
\end{eqnarray}
\begin{figure} [h]
\includegraphics[scale=0.45]{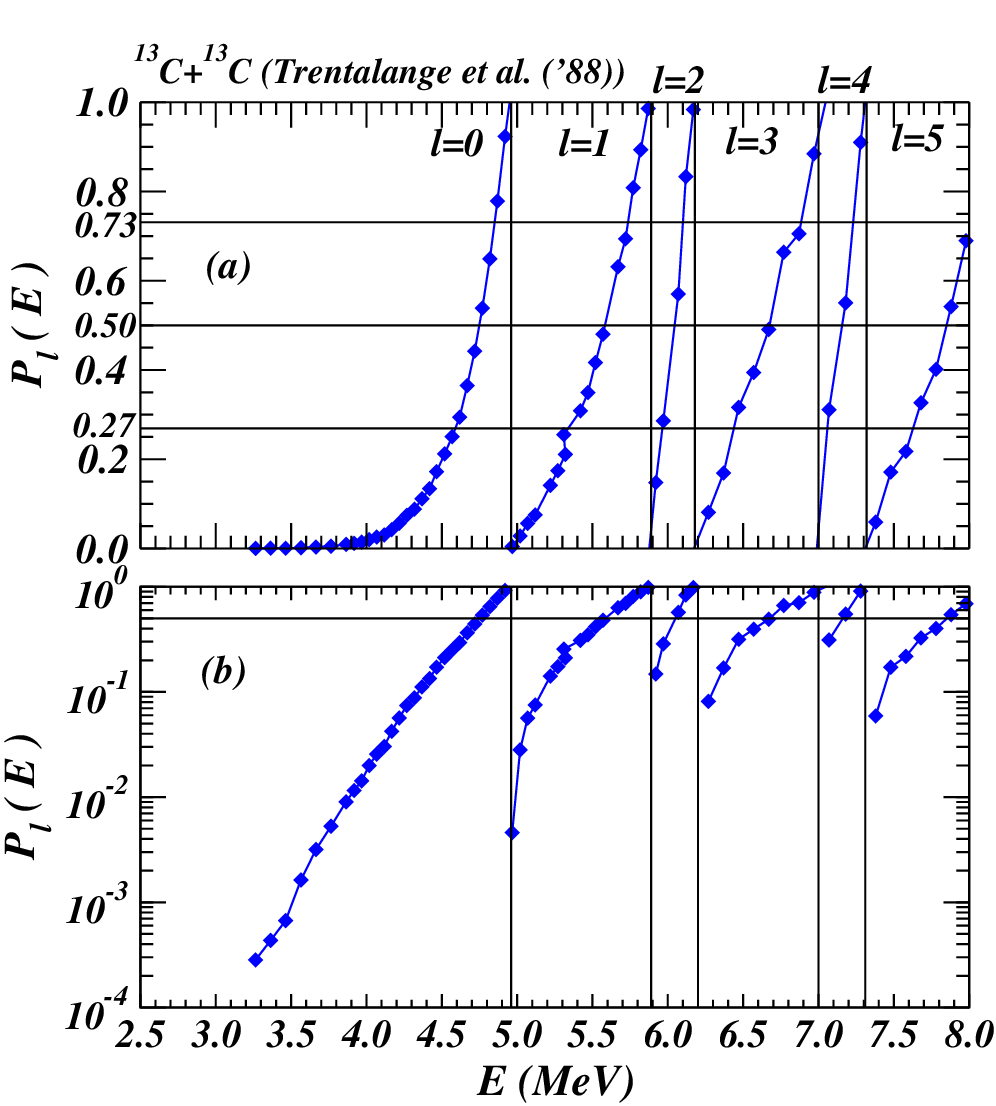}

\caption{(Color online) The penetration probability $P_l(E)$ as a
  function of $E$ on a linear scale (a) and on a logarithmic scale (b)
  extracted from the ${}^{13}$C+ ${}^{13}$C data of \cite{Tre88}.  }
\end{figure}

Under the assumption that $|E_l-E_j| \gg \Gamma_l$ for $l\ne j$
 for light nuclei collisions, the
contributions of different partial waves to the dimensionless cross
section are well separated in energy.  In the collision of identical
spin-1/2 nuclei , the penetration probability for the even-$l$ partial
wave is given by
\begin{eqnarray}
P_l(E) = \frac{\Sigma(E)-B(l)}{(2l+1)/2} \Theta[T(l)-\Sigma(E)]\Theta[\Sigma(E)-B(l)],\nonumber
\\ \label{cc}
\end{eqnarray}
where  $T(l)$ with even $l$ is given by  
\begin{eqnarray}
T(l)&=&l(l+1)+(l+1)/2,
\end{eqnarray}
and the function $B(l)$ is given by
\begin{eqnarray}
B(0)&=&0  ~~~ {\rm and~~} B(l)=T(l-1)~~{\rm for~~} l\ge 1 .
\end{eqnarray}
In the collision of identical spin-1/2 nuclei, the penetration
probability for the odd-$l$ partial wave is
\begin{eqnarray}
P_l(E) = \frac{\Sigma(E)-B(l)}{3(2l+1)/2} \Theta[T(l)-\Sigma(E)]\Theta[\Sigma(E)-B(l)],\nonumber
\\ \label{dd}
\end{eqnarray}
where $T(l)$ with 
odd $l$ is
\begin{eqnarray}
T(l)&=&l(l+1) +(3l+3)/2, 
\end{eqnarray}
and the function $B(l)$ is given again by
\begin{eqnarray}
B(0)&=&0  ~~~ {\rm and~~} B(l)=T(l-1)~~{\rm for~~} l\ge 1 .
\end{eqnarray}

Using Eq.\ (\ref{cc}) or (\ref{dd}) and the ${}^{13}$C+ ${}^{13}$C
data from \cite{Tre88}, we extract the penetration probability
$P_l(E)$ as a function of $E$ for various $l$ partial waves.  The
results are shown on a linear scale in Fig.\ 8(a) and on a logarithmic
scale in Fig.\ 8(b).  In Table III, we list the fusion barrier $E_l$
and the width $\Gamma_l$ extracted from $P_l(E)$ in such a procedure
for ${}^{13}$C+ ${}^{13}$C.  As one observes, the widths for the
even-$l$ states and the widths from the odd-$l$ states appear to fall
into two different groups, with $\Gamma$ for the even-$l$ states in
the 0.7-0.12 MeV range, while the width parameters $\Gamma$ for the
odd-$l$ states lie in the 0.19-0.22 MeV range.  There seems to be strong
dependence on the  even or odd property of the angular momentum $l$ of the
fused system.

\begin{table}[h]
 \caption {  The empirical values of $E_l$ and $\Gamma_l$  from  $P_l(E)$,  as extracted from the data of \cite{Tre88} for the collision of ${}^{13}$C+ ${}^{13}$C.
}
\vspace*{0.2cm} 
\hspace*{0.0cm}
\begin{tabular}{|c|c|c|c|c|c|c|c|c|}
\hline    
  $l$   & 0 & 1 & 2 & 3 & 4 & 5   \\\hline    
$E_l$ (MeV)  &~ 4.70~ & ~5.59~ &~ 6.04~ & ~6.68~ &~7.16~ & ~7.86~  \\
\hline
$\Gamma_l$ (MeV)  &~ 0.12~~ & ~0.19~~ &~~ 0.07~~ & ~0.22~ & ~0.10~~ & ~0.20~ \\
\hline
\end{tabular}
\end{table}

It should be noted that in deriving the barrier and penetration
probability rules, we have made the assumption that $|E_l-E_j|\gg
\Gamma_l$ for $l\ne j$.  The results in Table II and III for $E_l$ and
$\Gamma_L$ indicate that such an assumption is substantially valid and
is a reasonable and approximate idealization for light nuclei collisions . The extracted barrier
height and penetration probabilities are approximate quantities that
reveal the gross features of the fusion process.

\section{Conclusions and Discussion}

By treating the angular momentum as a continuous variable, the
reaction cross section can be evaluated in a simple analytical form.
The continuum approximation of the discrete angular momentum variable
carries errors, and these errors can be evaluated and amended to
previous results.

Three different formulas have been presented in the present formulations.
Formula I corresponds to the earlier result of
Ref.\ \cite{Won73} using the continuum approximation for all
partial waves.  Formula II is obtained by writing out the contribution
from the lowest $l=0$ partial wave explicitly and treating the higher $l\ge 1$
partial wave contributions in the continuum
approximation. Formula III involves Formula II with the
inclusion of corrections up to the second order in $\Delta l = 1/2$.

For the collision of unequal nuclei, the better formula is Formula II,
which incurs errors of order 0.7\% in the sub-barrier regions and
errors of order 0.2\% at high energies.  The simpler Formula I
incurs errors of about 4.4\% in the sub-barrier region, and  errors
of about 0.4\% at high energies.  Higher order corrections in Formula
III can be used if high accuracy is desired, with errors of about 0.005\%
in the sub-barrier region, and errors of about 0.12\% at high
energies.

For the collision of identical spin-0 nuclei, the application of these
formulas incur substantial errors.  The best formula for identical
spin-0 nuclei is Formula II, which incurs errors about 0.2\% in the
sub-barrier regions and errors of about 5.0\% at high energies.  On
the other hand, the simpler Formula I incurs errors of about 6.0\% in
the sub-barrier region, and an errors of about 5.5\% at high energies.

Simple rules have been presented to determine the barriers
$E_l$ and the penetration probabilities $P_l(E)$ for different $l$
partial waves from experimental data, for the collision of identical
or non-identical light nuclei.  The direct determination of the physical
quantities as a function of $l$ gives new insight in the fusion
process.  The barrier analysis rule has been successfully
applied to examine the relation between the fusion barrier and
$l$ for the pathological case of ${}^{12}$C+${}^{13}$C.  The
application of the penetration probability analysis reveals
quantitatively the resonance structure in ${}^{12}$C+${}^{12}$C
collisions.

We note that the partitioning of the partial waves into the lowest $l$
region and the higher $l$ region has some advantages in phenomenology.
There are situations in which the properties of the potential barriers
for the lowest $l$ states may deviate from the systematics of those
for the higher $l$ states.  These lowest $l$ states may need to be
specially handled.  One may provide a different description of the
penetration probabilities for the lowest partial waves, with
contributions from higher $l$ partial waves represented analytically
in the continuum approximation.  By this partition, the new degrees of
freedom, if any, can be incorporated into the penetration probability
to provide a clearer picture of the dynamics of the fusion process.

For simplicity, we have carried out the barrier analysis and the
penetration probability analysis for light nuclei collisions.  For
collision with heavy nuclei, however, $\Gamma$ is not small compared
to adjacent barrier separations $|E_l-E_{l\pm1}|$.  The barrier analysis for low-$l$
partial waves needs to be carried out iteratively.  While analytical
expressions have been obtained to carry out such an iterative
procedure, whether such a barrier analysis for heavy-nuclei collisions
may be practical remains to be investigated.

\vspace*{0.3cm}

\centerline{\bf Acknowledgment}

\vspace*{0.2cm} The author acknowledges the benefits of tutorials at Princeton University from the late
Prof. John A. Wheeler, whose Hill-Wheeler penetrability formula laid the foundation for the present work.
The author wishes to thank Prof.\ Xiao-Dong Tang for
stimulating discussions and helpful communications.  This research was
supported in part by the Division of Nuclear Physics, U.S. Department
of Energy.

\end{document}